\documentstyle[preprint,aps,psfig,array,floats]{revtex}
\ifx\undefined\psfig\def\psfig#1{ }\else\fi

\def\be{\begin{equation}}
\def\ee{\end{equation}}
\def\bea{\begin{eqnarray}}
\def\eea{\end{eqnarray}}
\def\ba{\begin{array}}
\def\ea{\end{array}}

\begin{document}
\ifpreprintsty\else
\twocolumn[\hsize\textwidth%
\columnwidth\hsize\csname@twocolumnfalse\endcsname
\fi

\preprint{IPM-00-051}
\title{{\large\bf THE COMPOSITION LINES OF THE VISIBLE BAND OF SYNTHETIC DIAMOND}}
\vspace{0.5cm}
\author{
{\bf M.A.
Vesaghi${}^a$\footnote{E-mail:vesaghi@sina.sharif.ac.ir},
 A. Shafiekhani${}^b$\footnote{E-mail: ashafie@theory.ipm.ac.ir} and K. Horiuchi${}^c$}}
\address{${}^a$ Dept. of Physics, Sharif University of Technology,\\
P.O.Box: 11365-9161, Tehran, Iran\\
${}^b$ Institute for Studies in Theoretical Physics and Mathematics,\\
P.O.Box: 19395-5531, Tehran, Iran\\
${}^c$ Frontier Technology Research Institute, Tokyo Gas Co. Ltd.,\\
1-7-7 Suehiro-cho, Tsurumi-ku, Yokohama 230, Japan\\
}

\maketitle
\begin{abstract}
\leftskip 2cm \rightskip 2cm The visible band (band A) of
Photoluminescence spectra of high-purity synthesized diamond is
analyzed by deconvolution technique. A set of eight lines with distinct
peak energy are found. The peak energy and the width of these lines were
either constant or varies very slightly with temperature. The amplitude of
the lines are significantly temperature. The
closeness of the temperature that the amplitudes of these lines
reach their minimum and the temperature which the Free-exiton
emission is maximum at, is an indication of the competition
between these too effects.
\end{abstract}
\ifpreprintsty\else\vskip1pc]\fi
\newpage
The vacancies and their related properties in diamond and
Diamond-Like Carbon (DLC) both in theoretical aspect\cite{vs1} and
experimental results\cite{vs2} are fascinating subjects. DLC
produced by different techniques, artificial diamond produced by
high pressure and high temperature show peculiar
Photoluminescence (PL)\cite{lwx,hnkk},
Cathodeluminescence (CL)\cite{ck1} and absorption\cite{vs2}. There
are different peaks and cranks in the absorption, PL and CL
spectra of these materials, which some of them are related to the
vacancies present in them. Many of these peaks are not simple and
are a combination of different absorption or emission
lines\cite{vs2}. Undisturbed vacancies have almost all the
rotational symmetry of the main crystal, which for diamond is
${\rm T}_d$. The zero phonon line at 1.673eV and the line at
3.145eV which reported by many groups, are associated to
undisturbed neutral vacancies (${\rm V}^\circ$)\cite{ck2} and
undisturbed negatively charged vacancies (${\rm V}^-$)\cite{lo1}
respectively. Much work has been done on the effect of uniaxial
stress in natural diamond to explain the lines of the general
radiation centers band (GR1) that has many peaks from 1.5eV up to
2.8eV. Strong internal stress exist in DLC films\cite{zsrd}.
It has been reported that for films there is a strong
correlation between PL intensity and stress\cite{fs}. In this
article, we show that for artificial diamond similar effect is
possible due to local stress effect.

The data analyzed here, is for the sample that was grown by high
pressure and high temperature-gradient method where
nitrogen-getter was added to the solvent. The nitrogen
concentration was estimated to be less than 0.1 ppm by
infra-red(IR) absorption spectroscopy. The full width at half
maximum of diamond Raman peak at 1333 ${\rm cm^{-1}}$ is 1.6 ${\rm
cm^{-1}}$ and smaller than that of typical natural type IIa
diamond, 1.9 ${\rm cm^{-1}}$, which indicates a low concentration
of defects in the sample. Still the effect of these defects
(vacancies surrounding metallic atoms) is small but visible in
Photoluminaces spectra.

The PL spectra of this crystal at several temperatures from
$400^\circ$K down to $8.4^\circ$k were measured and reported by K.
Horiuchi and his coworkers\cite{hnkk}. The complete spectra at
room temperature is shown in Fig. (1).
\begin{figure}
\centerline{\psfig{figure=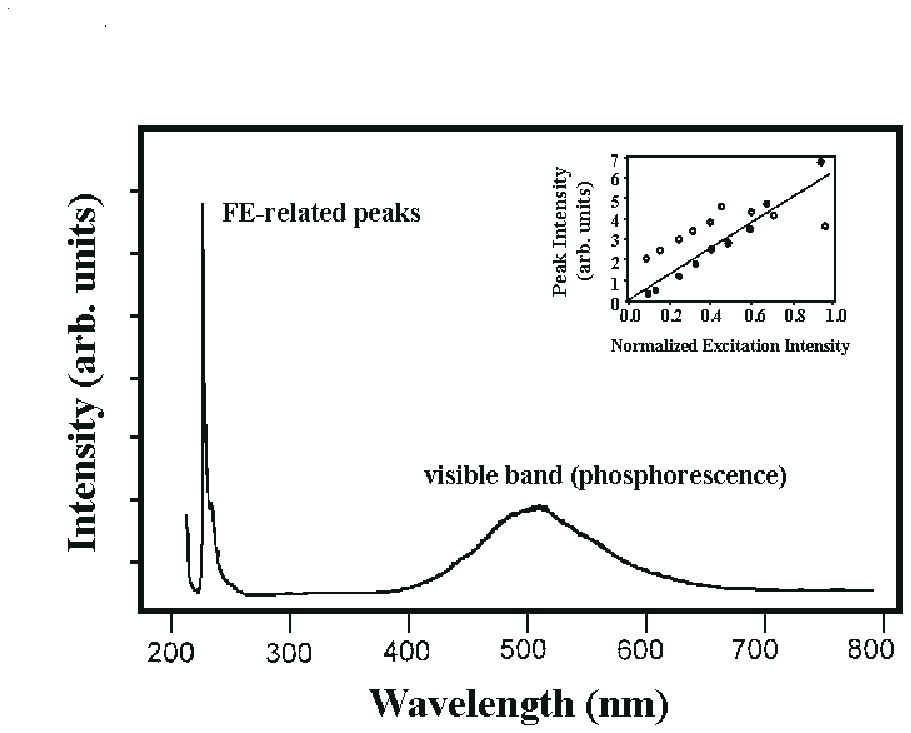,width=.45\columnwidth}}
\vspace{5mm} \caption{PL spectra at room temperature, corrected
for the measurement sensitivity. The shoulder at 220 nm is due to
Rayleigh scattering and Raman peak of diamond. The inset shows the
peak intensity dependence on input power; open circle, visible
band (at 520 nm); close circle, FE emission (at 235 nm). The line
is drawn to guide the eye.[4]}
\end{figure}
 The strong exiton line in the room temperature spectra of
this crystal is also the evidence of its high quality\cite{hnkk}.
At all temperature the PL spectra of this sample has a broad and
complicated band from 1.5eV to 3.2eV, which are shown in Fig. (2).
In the paper by Horiuchi {\it et all} was written: "the peaks from
350 nm up to 750 nm (3.5 eV down to 1.65 eV) are the combination
of two bands one at 570 nm (2.17 eV) and the second one around 520
nm (2.38 eV)". As it is apparent from the spectra and through our
experience, there are more than two lines present in these
spectra. We used the deconvolution technique to analyze these
reported results.
\begin{figure}
\centerline{\psfig{figure=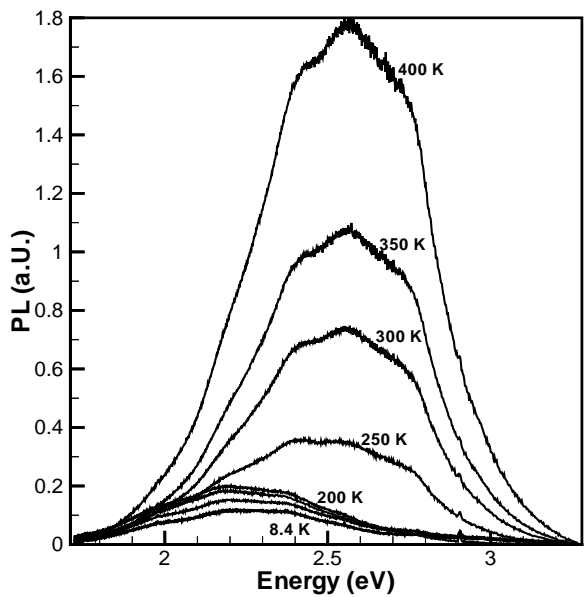,width=.45\columnwidth}}\hspace{1cm}
\vspace{5mm} \caption{ PL spectra of perfect diamond crystals
grown by high pressure and high temperature technique, at
different temperatures from $400^\circ$K down to $8.4^\circ$K. }
\end{figure}

The spectra obtained at different temperatures were deconvoluted
by normalized Gausians with a greater than 99.95$\%$ accuracy for
high temperature ones down to 99.5$\%$ accuracy for low
temperature ones. We used as guide for our deconvolution, the
cranks in the spectra and the narrow lines that were nicely
apparent at low temperature. All spectra are deconvoluted to eight
normalized Gaussian  lines. The results for $400^\circ$K and
$8.4^\circ$K are shown in Fig. (3). For other temperatures the
results are similar and the pick energy of each line remain
constant. At each temperature the widths of all lines are the same
and changes slightly as the temperature is lowered. The amplitude of
each lines varies significantly with temperature.
The deconvolution results are given in table (1).

 Fig.(4) shows the temperature
dependence of the amplitude of these lines. The amplitude of all
the lines decrease as the temperature is lowered and all have
minimums at some temperatures in the range of $150-200^\circ$K.
The temperatures that these lines have minimum at, are close to
the temperature where the amplitude of Free-exciton (FE) related
emission is maximum\cite{hnkk}.
These two, emission via
FE and emission via vacancies are competing in this temperature region.
In addition to the
explanation of Horiuchi {\it etal.} for the existence of a trap,
one might say that some of these lines are related to each other
and to neutral vacancy $V^\circ$ via Jahn-Teller effect\cite{vs2}.
The ground state of $V^\circ$ is doubly degenerated and close to
the top of the valance band\cite{lo2}. Local strong stress exists in
this artificial diamond due to the metallic atoms present in it.
Because of this stress, there would be a symmetry braking. The
symmetry braking splits the degenerate states and as a consequence
the $V^\circ$ ground state will split to two sub-levels, one very
close to the top of the valance band and the second above it. The
state close to the top of the valance band interferes with (FE)
formation. With temperature decrease this interference reduces and
the (FE) effect increases. For $V^-$ similar effect could be happened.
The large width of these lines are due to the size of the clusters
surrounding each metallic atom.

By deconvolution of the Photoluminescence spectra of artificial
diamond at different temperatures we found that as it is in the
case of DLC films, there exist vacancies under stress in the regions
surrounding metallic atoms. In addition to
this there might be a competition between Photoluminescence via
vacancies and via (FE).

\begin{figure}
\centerline{\psfig{figure=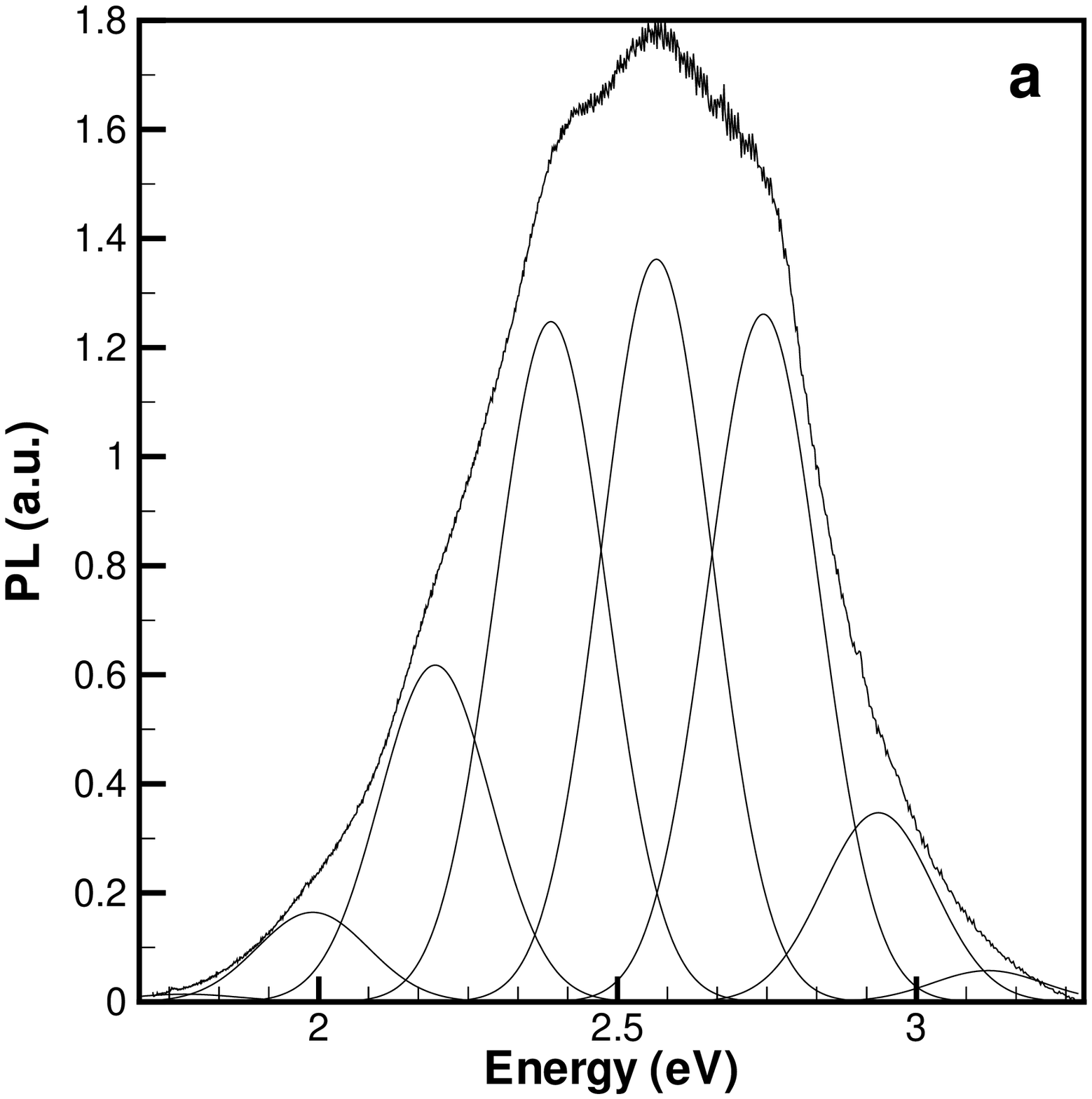,width=.35\columnwidth}\hspace{.8cm}
\psfig{figure=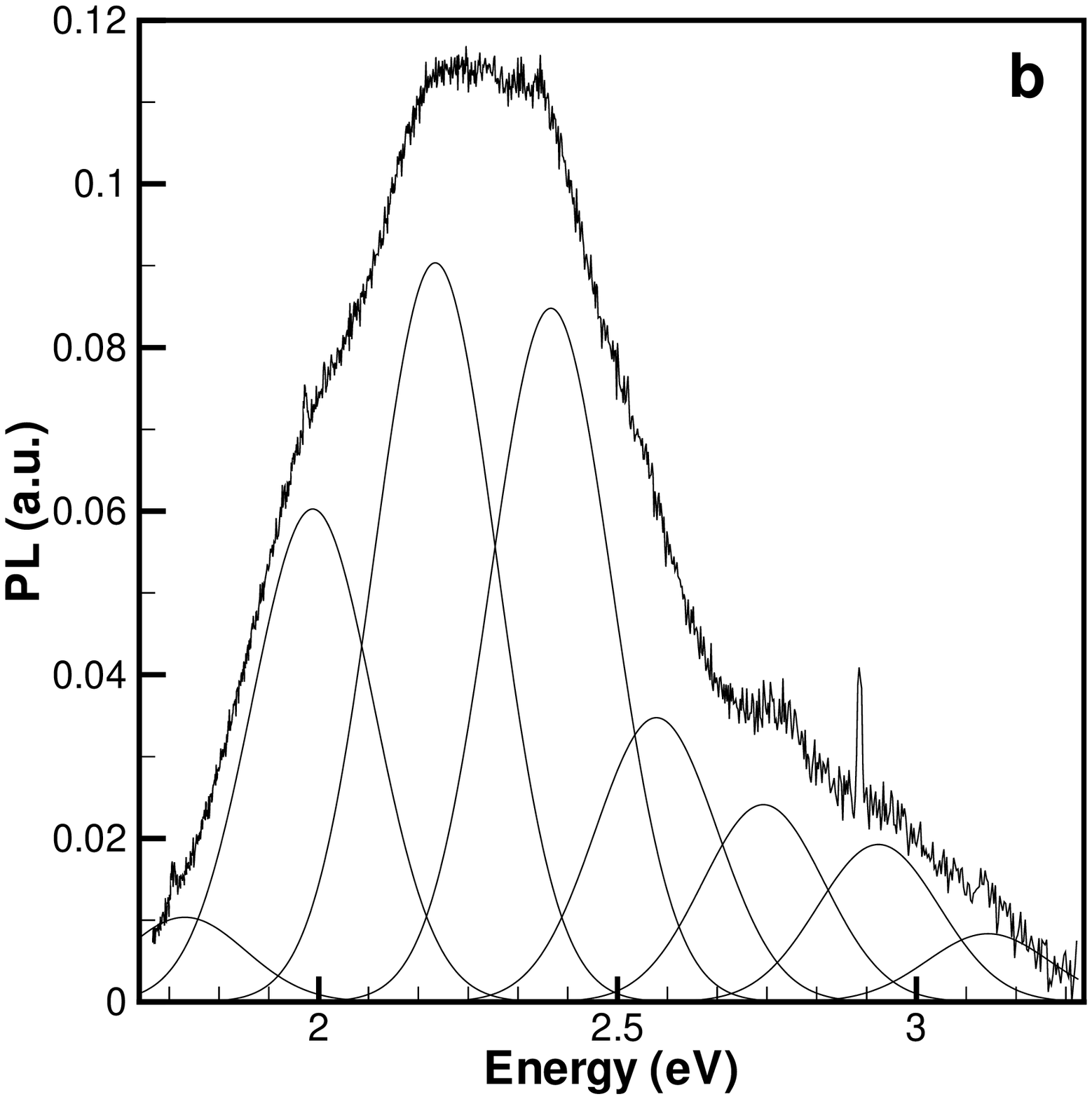,width=.35\columnwidth}} \vspace{5mm}
\caption{Deconvoluted Photoluminescence spectra of artificial diamond
taken at a) 400${}^\circ$K and b) 8.4${}^\circ$K. At each temperature
the width of all Gussian lines are the same.}
\end{figure}
\begin{center}
\begin{table}
\caption{The deconvolution results of Photoluminescence spectra of
artificial diamond. First row is the temperature, T(K), where each
spectra were taken at. The tenth row is the width, W, of relative
peak in eV$\times 10^{-5}$. The last row is the deconvolution
accuracy, Acc.(\%). The first column is the peak energy, E(eV).
The rest in the entire of the table is amplitude in arbitrary unit
} \vspace{0.5cm}

 \scriptsize
\begin{center}
\begin{tabular}{|c||c|c|c|c|c|c|c|c|c|}
    \hline
    ${{\rm T(K)}_\rightarrow}$&400&350&300&250&200&150&100&50&8.4\\ \cline{1-1}
E(eV)$\downarrow$&&&&&&&&&\\ \hline\hline
1.776&187&159&230&306&421&345&227&170&149\\ \hline
1.990&2156&1380&1220&1263&1500&1600&1147&850&867\\ \hline
2.195&8100&4960&3680&2600&2139&2330&1769&1360&1300\\ \hline
2.389&16360&9920&6960&3740&1710&1870&1496&1220&1220\\ \hline
2.565&17860&10810&7420&3460&750&794&661&510&500\\ \hline
2.744&16540&9700&6390&2790&734&454&454&320&347 \\ \hline
2.937&4550&2490&1550&600&47&236&292&237&277\\ \hline
3.112&754&352&210&67&40&90&133&85&120\\ \hline\hline
W(eV)&1720&1720&1720&1760&2070&2100&2062&2030&2070\\ \hline\hline
Acc.(\%)&99.956&99.958&99.957&99.93&99.88&99.85&99.77&99.676&99.659\\
\hline
\end{tabular}
\end{center}
\end{table}
\end{center}
\vspace{5mm}
\begin{figure}\label{4}
\centerline{\psfig{figure=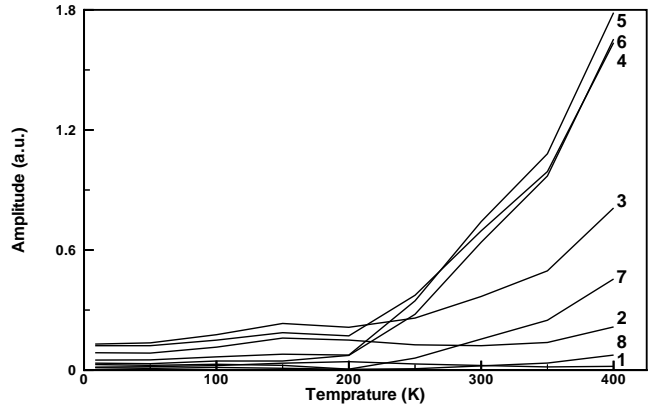,width=.5\columnwidth}\hspace{.8cm}}
\caption{ The temperature dependence of the amplitude of spectra
lines}
\end{figure}
\normalsize
 
\end{document}